\documentclass{Rinton-P9x6}

\begin{document}

\title{Solar and atmospheric neutrinos: a solution with large extra dimensions}

\author{Dominique Monderen}

\address{ULB Service de Physique Th\'eorique \\
Campus Plaine CP225 \\
Bld. du Triomphe, 1050 Brussels, Belgium \\
E-mail: dmondere@ulb.ac.be
}

\maketitle

\abstracts{
We have studied \cite{nous} flavor neutrinos confined to our four-dimensional world coupled
to one "bulk" state, i.e., a Kaluza-Klein tower. Within a minimal model, we accomodate
existing experimental constraints (CHOOZ experiment, solar 
and atmospheric data). No direct flavor oscillations nor MWS effects are needed.
An energy independent suppression is produced at large distances.}
\section{Introduction}
The possibility of "large extra dimensions", i.e., with (at least) one
compactification radius close to the current validity limit of Newton's law of gravitation
($\sim$ 0.1 mm), raises considerable interest.
It could also solve the hierarchy problem by providing us with a new fundamental scale
at an energy possibly as small as 1 TeV \cite{led}. 

Neutrino physics is a favored area to study this possibility. A right-handed, sterile neutrino
does not experience any of the gauge interactions and is therefore not confined to our
four-dimensional brane; it is thus, other than the graviton, an ideal tool to probe the "bulk" of
space. The bulk states appear to us (due to compactification of the extra
dimensions) as so-called Kaluza-Klein towers of states \cite{lednu}. 
Recent works \cite{dvali,lorenzana,creminelli,nuled} have shown that  
it is, at least partially, possible to accommodate experimental constraints 
on neutrinos within this setup.

We explore further possibilities, focusing on the unique properties of the
model. Quite specifically, neutrinos in this scheme can "escape" for part of the time to extra dimensions,
resulting in a reduced average probability of detection in our world. While similar in a way to a fast unresolved
oscillation between flavor and sterile neutrinos, this differs both by the time development of the effect and
by the possible depth of the suppression.
\section{Two neutrinos coupled to one bulk fermion: the 2-1 model}
The simplest model is constituted by one left-handed neutrino state $\nu _1$
which lives in our 3+1 dimensional world coupled with one singlet ''bulk'' massless fermion
field. Since the latter lives in all dimensions, from our world's point of view, it
appears, after compactification, as a Kaluza-Klein (KK) tower, i.e., an infinite
number of four-dimensional spinors.

The analysis presented here is based on a reduction of the theory from 4+1
to 3+1 dimensions. However, to guarantee a low scale for the unification of
gravity with all forces, more extra dimensions are needed. We will assume
that their compactification radii are small enough that they don't affect
the analysis. The pattern is now well established (see e.g.~\cite{dvali,creminelli})
and we will only recall the basic equations and results.
The action used is the following : 
\begin{equation}
\label{action}
S=\int d^4x\,dy\;\overline{\Psi} i\Gamma _A\partial ^A\Psi +\int d^4x\{\overline{\nu} _1 
i\gamma _\mu \partial ^\mu \nu _1+\lambda \overline{\nu} _1\Psi (x^\mu ,y=0)H(x^\mu
)+h.c.\} 
\end{equation}
where $A=0,...,4$ and $x^4=y$ is the extra dimension. 
The Yukawa coupling between the usual Higgs scalar, 
the weak eigenstate neutrino $\nu _1$ and the bulk
fermion operates at $y=0$, which is the 3+1
dimensional brane of our world.

The fifth dimension is taken to be a circle of radius $R$. As usual, the
bulk fermion $\Psi $ is expanded in eigenmodes. 
One then integrates over the fifth dimension.
Eventually, one has to diagonalize the mass matrix (eigenvalues noted 
$\lambda _n$) and write the neutrino in terms of the mass eigenstates : 
\begin{eqnarray}
\label{valprop}
\lambda _n =\pi \xi ^2\cot (\pi \;\lambda _n) \;\; {\rm ;} \;
\left| \nu _1\right\rangle =\sum _{n=0}^{\infty }U_{0n}\left| \nu
_{\lambda _n}\right\rangle \;\; {\rm ;} \;
U_{0n} ^2 =\frac 2{1+\pi^2 \xi ^2+ \frac{\lambda _n ^2}{\xi ^2}}
\end{eqnarray}
where $\xi \equiv \frac m{1/R}$ 
measures the strength of the Yukawa coupling\footnote{another convention introduces a 
$\sqrt{2}$ factor, as in~\cite{dvali,lorenzana}} .

The survival amplitude $A_{\nu _1\nu _1}$and probability 
$P_{\nu _1\nu _1}$ are given by 
\begin{eqnarray}
\label{anu1nu1}
A_{\nu _1\nu _1} &=&\sum _{n=0}^{\infty }U_{0n}
^2e^{\,i \,\lambda _n ^2x} \\
\label{pnu1nu1}
P_{\nu _1\nu _1} &=&\sum _{n=0}^{\infty } U_{0n} ^4+\sum
\sum _{n \neq m} U_{0n} ^2 U_{0m} ^2\cos
\left[  \left( \lambda _n ^2-\lambda _m
^2\right) x\right] \\
{\rm where} \;\;\;
x &=& \frac{L}{2ER^2} \approx 10^{-7}
\frac{\left( L/ {\rm km} \right) }{(E/ {\rm GeV})(R/ {\rm mm})^2} \nonumber
\end{eqnarray}
We have shown in Ref.\cite{nous} that for $\nu_1 = \nu_e$, it is possible
to accomadate the CHOOZ contraint with a global $L/E$ independent solar suppression
(free of MSW effect) but an atmospheric $\nu_e$ suppression is then expected.
To reconcile with the data in that case, one needs the MSW effect \cite{dvali,creminelli,nuled}.
To keep the solar suppression $L/E$ independent, we do not further investigate this
possibility.
$\nu_1$ can however be a linear combination of flavor states. Indeed if we describe the coupling
of two\footnote{It is also possible to replace $\nu_\mu$ by $\nu_\tau$ or even to introduce the
three flavor states, as in \cite{dienes} or to couple the bulk to a four-dimensional sterile state.}
flavor neutrinos to the bulk neutrino by the Lagrangian
\begin{equation}
\label{lagrangian}
{\cal L} =\lambda _{e}\overline{\nu }_{e}\Psi (x^{\mu },y=0)H(x^{\mu })
+\lambda _{\mu }\overline{\nu }_{\mu }\Psi (x^{\mu },y=0)H(x^{\mu }) {\rm ,}
\end{equation}
we recover the previous action (\ref{action}) by a rotation of the flavor base
\begin{eqnarray}
\nu _{1} &=&\cos \theta \;\nu _{e}+\sin \theta \;\nu _{\mu } \cr
\nu _{2} &=&-\sin \theta \;\nu _{e}+\cos \theta \;\nu _{\mu } {\rm ,}
\end{eqnarray}
with $\cos \theta =\frac{m_{e}}{m}$ , $\sin \theta =\frac{m_{\mu }}{m}$, $m=%
\sqrt{m_{e}^{2}+m_{\mu }^{2}}$ and $m_{e,\mu }=\frac{\lambda _{e,\mu}v}{%
\sqrt{2\pi R}}$ \footnote{Here, $m_{e}$ and $m_{\mu}$ are simply mass parameters
without any link to the charged fermion masses.}.
The mixing with bulk states remains unchanged for $\nu_1$.
The orthogonal combination $\nu _{2}$ remains
massless and decouples from the bulk neutrino.
A new phenomenological parameter, 
the mixing angle $\theta $ now plays a crucial role
in the survival probabilities of the flavor neutrinos,
\begin{eqnarray}
P_{\nu _{e}\nu _{e}} &=&\cos ^{4}\theta \;P_{\nu _{1}\nu _{1}}+\sin
^{4}\theta +2\sin ^{2}\theta \;\cos ^{2}\theta \;\textrm{Re}\left( A_{\nu
_{1}\nu _{1}}\right) \cr
P_{\nu _{\mu}\nu _{\mu }} &=&\sin ^{4}\theta \;P_{\nu _{1}\nu _{1}}+\cos
^{4}\theta +2\sin ^{2}\theta \;\cos ^{2}\theta \;\textrm{Re}\left( A_{\nu
_{1}\nu _{1}}\right)
\end{eqnarray}
and a flavor transition is possible through the bulk states
\begin{equation}
P_{\nu _{e}\nu _{\mu }}=P_{\nu _{\mu}\nu _{e}}=\sin ^{2}\theta \ \cos
^{2}\theta \ (P_{\nu _{1}\nu _{1}}-2\textrm{Re}\left( A_{\nu _{1}\nu
_{1}}\right) +1) 
\end{equation}

This model (hereafter called the 2-1 model) has 3 degrees of freedom, 
$(\xi ,\theta ,R)$, to fit the
experimental data.
We should also stress that $\theta$ is not the $\nu _e - \nu _\mu$ mixing angle
\footnote{The usual mixing angle is a rotation from the flavor eigenstates to
mass eigenstates. Here $\theta$ describes a rotation to the eigenstates of the coupling
(\ref{lagrangian})}. Such
mixing  arises, but only as a result of the independent coupling of both states to the bulk
neutrino.
\subsection{Confrontation of the model with experimental data}
We discuss here how the experimental constraints can be accounted for in our 2-1 model.
A discussion of the experimental data themselves and related references can be found in
Ref.\cite{nous}.

The solar neutrino deficit is accounted for if the $\nu _{e}$ mean survival 
probability $\left< P_{\nu_e \nu_e} \right>$ at large $x$ ranges between 40\% to 60\% (as a 
result of the experimental errors and solar model uncertainties).
This constraint defines a region in the plane $\xi-\theta$ (Fig.~6 in Ref.\cite{nous}).
In models with large extra dimensions and if no MSW effects are playing,
$\left< P_{\nu_e \nu_e} \right>$ is naturally constant at large $x$.
Such a solution is the simplest interpretation\footnote{
The situation is similar to vacuum oscillations at large $\delta m^2$ but
there the suppression cannot exceed 50\%.} of the absence 
of $L/E$ dependence in the SuperKamiokande solar neutrinos data.
To prevent any $L/E$ dependence, we should also avoid MSW effects in the Sun or Earth.

As the Sun-Earth system is a very long-baseline one and as the solar core is large 
(typically, $x \sim 10^5$ and $\Delta x \sim 10^2 \gg 1$ for solar neutrinos 
with $E \sim 1$ MeV and $R \sim 1$ mm),
$\left< P_{\nu_e \nu_e} \right>$ is the only observable effect:
the fast fluctuations of the survival probability
are completely washed away, whatever the detector resolution.

The CHOOZ nuclear reactor experiment observes no $\overline{\nu }_{e}$ disappearance
at a distance $L=1~$ km and a typical energy of 2 MeV. 
For given $\xi$ and $\theta$, a maximum admissible value of $x$,
or equivalently, a minimum value 
of $R$ (the radius of the compactified extra dimension) results.
A small coupling constant $\xi$ and 
a large mixing angle $\theta$ are preferred.

On the other hand, $1/R$ controls the typical mass difference between two
consecutive Kaluza-Klein levels. Therefore, MSW resonant conversion will
take place if $1/R$ is of the same magnitude order as the MSW potential. 
To avoid the MSW effect and consequently $L/E$ dependences in the mean survival
probability, we can put an upper bound on $R$.
Typically, $R_{\max }\simeq 10^{-2}$ mm. As a result, for some $\xi$ and $\theta$,
this bound can be incompatible with the CHOOZ constraint (see Fig.~6 in Ref.\cite{nous}).

The $\overline{\nu }_{\mu }$ disappearance experiment K2K reveals some 30\% deficit for 2 GeV
neutrinos at a distance $L \simeq 250$ km. Since we have $x_{\rm K2K}\simeq 1/4 \, x_{\rm CHOOZ}$,
$\nu_\mu$'s are expected to disappear more than $\nu_e$'s. This requires
$\theta >\pi /4$, and higher values of $\xi $ are favored, as $P_{\nu _{e}\nu _{e}}-P_{\nu
_{\mu }\nu _{\mu }}\propto (1-P_{\nu _{1}\nu _{1}})$. However, even for the
maximal allowed $\xi$, the preliminary result of K2K can only be accommodated by taking
the large statistical error into account. The Fig.~1 shows a possible fit for $\nu _e$ 
and $\nu _\mu$, which solves the solar neutrino problem, and simultaneously
satisfies the CHOOZ and K2K constraints.
\begin{center}
\begin{figure}[t]
\epsfxsize=27pc
\epsfbox{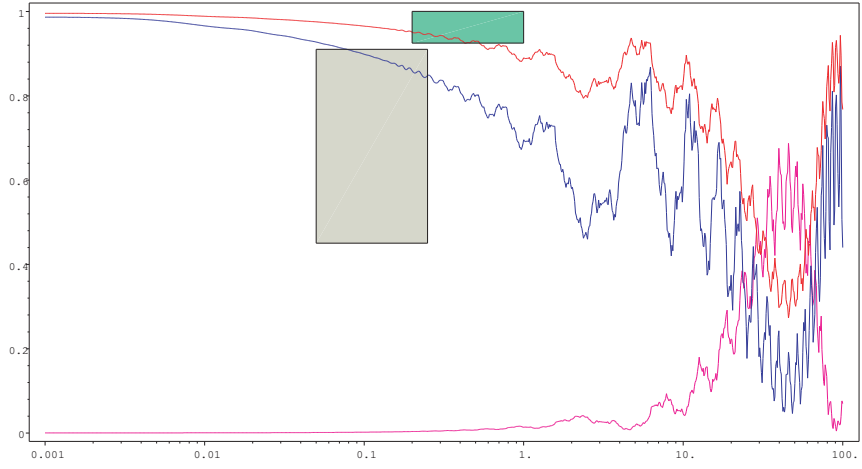}
\vspace{-5mm}
\caption{Comparison of the 2-1 model with the CHOOZ and K2K constraints (highest curve
is for $\nu_e$. $\xi=0.3$ and $\theta=1.05$, so that $\left< P_{\nu_e \nu_e} \right> \simeq 60\%$.
The transition probability $P_{\nu_e \nu_\mu}$ is also depicted (lowest curve). The error bar
(at 2$\sigma$ for CHOOZ; 1$\sigma$ for K2K) combines quadratically the statistical and systematic
errors.}
\vspace{-8mm}
\end{figure}
\end{center}
\begin{center}
\begin{figure}[t]
\epsfxsize=27pc
\epsfbox{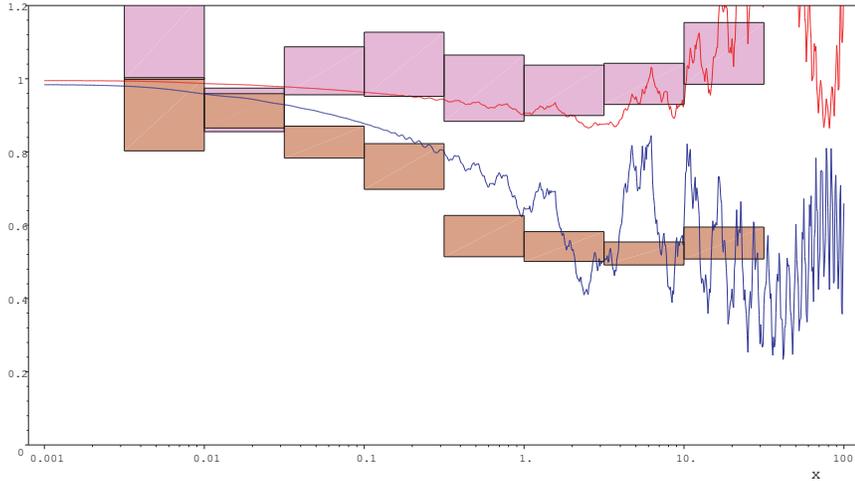}
\vspace{-5mm}
\caption{Expected atmospheric neutrino fluxes in the 2-1 model and the
SuperKamiokande data. Same $\xi$ and $\theta$ have been used. The series of filled boxes show the
$L/E$ dependence for the atmospheric $\nu_e$ and $\nu_\mu$, as observed by SuperKamiokande
(1$\sigma$). The initial flux is normalized to 1 at $x=0$ and the data has been normalized by an
overall 0.95 factor with respect to the raw data (instead of the usual $\sim$0.9 used by
SuperKamiokande). The agreement with experimental data is quite remarkable. The boxes of both
figures are taken as in $^1$.}
\vspace{-8mm}
\end{figure}
\end{center}

\vspace{-10mm}
To discuss the atmospheric neutrinos, we recall that in the 
2-1 model, a transition $\nu _e \rightarrow \nu _\mu$ or $\nu _\mu \rightarrow \nu _e$ 
becomes possible. The transition probability, as shown in Fig.~1, is non-negligible
in the range of the atmospheric neutrinos. As the latter originate from the 
decay of the charged pions and kaons into muons and the subsequent decay of muons,
the ratio of the neutrino initial fluxes 
$\phi ^{(i)}_{\nu _\mu} / \phi ^{(i)}_{\nu _e}$ 
is expected to be very close to 2, especially at low energy\footnote{At higher
energy, the produced muon can go through the atmosphere without decaying, so that the
ratio $\phi ^{(i)}_{\nu _\mu} / \phi ^{(i)}_{\nu _e}$ increases with energy.}. 
Therefore, the expected neutrino flux in the 
2-1 model is given by 
$\phi _{\nu _e} / \phi ^{(i)}_{\nu _e}=P_{\nu _e \nu _e}+ 2 \, P_{\nu _\mu \nu _e}$
and
$\phi _{\nu_\mu} / \phi ^{(i)}_{\nu _\mu}=P_{\nu _\mu \nu _\mu} + 1/2 \, P_{\nu _e \nu _\mu}$
 (we don't distinguish between $\nu$ and $\overline{\nu}$).
As a result, the observed $\nu _e$ flux can be enhanced compared to the initial production 
flux. In Fig.~2, we see that this picture is in very good agreement with 
the SuperKamiokande results. Moreover models with extra dimensions escape at least partially
the SuperKamiokande constraints on sterile neutrinos. Indeed, while oscillations are blocked
by an MSW effect in the Earth in a $\nu_\mu$-$\nu_s$ maximal mixing picture, anti-neutrinos
always find a resonant KK state to oscillate to in the $\nu_\mu$-$\nu_{KK}$ model\cite{nous}.

We are left with the constraints of KARMEN and LSND. 
The negative result of the KARMEN experiment can easily be accommodated
but our model can never 
comply with the LSND results, for any allowed values of $\theta $ and $\xi $.
This however can be understood: the 2-1 model provides two mass scales,
$\lambda_0 / R \approx \xi / R =m $ and $\lambda_1 / R \approx 1/R$,
allowing to account for the solar and atmospheric anomalies.
The contributions of the KK tower can in principle account for the LSND results,
but it turns out that the effect is too weak.

We also point out 
that the astrophysical bound could be partly evaded in this model, since the disappearance 
of $\nu _e$ or $\nu _\mu$ in the extra dimensions is never complete (see~\cite{dienes}; for
discussions on astrophysical and cosmological constraints, see, e.g., \cite{creminelli,astro}).
\section{Conclusions} \label{s:concl}
We have shown that a simple model with 2 massless flavor neutrinos, namely $\nu_e$ and $\nu_\mu$,
coupled to  one Kaluza-Klein tower meets most experimental constraints (except for 
LSND), and differs from the oscillation image by the energy-dependence of 
the neutrino disappearance. This model can be developed by adding extra 
parameters in the form of bare masses for the neutrinos, while simply 
increasing the number of neutrinos coupled to the Kaluza-Klein states 
brings little gain.
\section*{Acknowledgements}
This talk is based on a work accepted for publication in Physical Review D \cite{nous}.

\noindent D.~M. benefits from a F.R.I.A. grant.

\end{document}